\documentclass[9pt, sigconf, nonacm]{acmart}
\makeatletter
\def\@mkbibcitation{}

\makeatother

\usepackage{amsmath} 
\usepackage{amsfonts} 
\usepackage[linesnumbered, ruled]{algorithm2e}
\usepackage[noend]{algorithmic}
\usepackage{xcolor}
\usepackage{booktabs} 
\usepackage{graphicx} 
\usepackage{svg} 
\usepackage{array} 
\usepackage{makecell}
\usepackage{tabularx} 
\usepackage{subcaption}
\usepackage{textcomp}
\usepackage{rotating} 
\usepackage{multirow}
\usepackage{diagbox}
\usepackage{caption}
\usepackage{footnote} 
\usepackage{fancyvrb} 
\usepackage{listings}
\usepackage{float}
\usepackage{pifont}

\newcolumntype{Y}{>{\centering\arraybackslash}X}

\newcommand{\ydel}[1]{}

\makeatletter
\def\@printendtopmatter{%
  \par
  \vskip 20\p@ 
  \@afterindentfalse
  \@afterheading
}
\makeatother

\begin{document}

\title{VClare: Resolving Imperfect Specifications in LLM-Based Verilog Generation}
\author{Zhuorui Zhao}
\affiliation{%
  \institution{Technical University of Munich}%
  \city{Munich}%
  \country{Germany}%
}

\author{Bing Li}
\affiliation{%
  \institution{Technical University of Ilmenau}%
  \city{Ilmenau}%
  \country{Germany}%
}

\author{Yu Li}
\affiliation{%
  \institution{Zhejiang University}%
  \city{Hangzhou}%
  \country{China}%
}

\author{Zheyu Yan}
\affiliation{%
  \institution{Zhejiang University}%
  \city{Hangzhou}%
  \country{China}%
}

\author{Ulf Schlichtmann}
\affiliation{%
  \institution{Technical University of Munich}%
  \city{Munich}%
  \country{Germany}%
}

\begin{abstract}
Large language models (LLMs) have demonstrated promising capabilities in generating Verilog code from natural language specifications.
However, human-written specifications often contain semantic imperfections such as vagueness, contradictions, and incompleteness, which can significantly degrade the quality of hardware design generated by LLMs.
In this paper, we present the first systematic study of
imperfect specifications and propose an automated framework {VClare} to repair them 
to enhance the quality of resulting Verilog design.
The proposed framework explores two complementary repair paradigms. The \textit{Spec-Level Repair} conducts LLM-driven inconsistency mining directly on the specification texts, while the 
\textit{Sim-Level Repair} employs simulation-based behavioral clustering 
with optional test-time inconsistency arbitration.
In addition, we propose two new benchmark datasets with systematically injected specification defects.
The first benchmark dataset is derived from the VerilogEval-human benchmark targeting single-module tasks, while the other benchmark dataset is derived from the ComplexVDB dataset and contains 53 multi-module tasks that reflect more realistic engineering scenarios.
For single-module tasks, the {VClare} framework can repair the imperfections in the specifications effectively and thus enhance the pass rate of the generated Verilog design by 12.7\%, while for the multi-module tasks this enhancement can reach 13.7\%, demonstrating the capabilities of specification repair by {VClare} as well as further potential of LLMs in front-end hardware design.\footnote{The two benchmark datasets are released at \url{https://anonymous.4open.science/r/VClare/}. The scripts will be open-sourced upon acceptance.}

\end{abstract}


\maketitle

\section{Introduction}
\label{sec:intro}

Large Language Models (LLMs) have demonstrated promising capabilities in generating Hardware Description Languages (HDL) such as Verilog directly from natural-language specifications~\cite{chang2024data, xu_kangwei_invited_2026}. By translating behavioral descriptions into HDL, LLMs offer the potential to accelerate hardware development and lower the barrier to digital design.
However, the correctness of the generated HDL fundamentally depends on the quality of the specification itself~\cite{lu_rtllm_2024}. Human-written specifications frequently contain semantic defects, including vagueness, contradictions, and incompleteness, especially in prototype specifications. Such defects can mislead LLMs into generating functionally incorrect hardware even when the downstream generation process is otherwise capable. Since specifications form the first stage of the hardware refinement pipeline, errors introduced at this stage will propagate throughout the design flow.

Existing work on LLM-based HDL generation largely assumes that specifications are semantically correct. They target improving LLM-generated HDL by repairing implementations after generation rather than repairing the specification itself.

\textit{Oracle-guided code repair:} Several approaches leverage external correcting signals such as iterative human conversation or golden testbenches. Chip-Chat~\cite{blocklove_chip-chat_2023} uses iterative human-LLM interaction to refine generated HDL. Later work~\cite{xu2024meic, thakur_autochip_2024, zhao2025mage, ho2025verilogcoder} automates this process using feedback-driven repair loops guided by golden testbenches. While effective, these approaches rely on strong external oracles that are expensive to obtain.

\textit{Oracle-free code repair:} Without an oracle, there is limited information available. Although LLMs can generate testbenches for HDL design \cite{qiu_autobench_2024, qiu2025correctbench}, the accuracy of such generated testbenches are the same, if not lower, than the accuracy of the generated hardware codes, making this an unreliable source of information. Therefore, early endeavors in the hardware domain focused on repairing syntactically incorrect code~\cite{tsai2024rtlfixer}, where compiler feedback alone suffices. More recently, methods such as \cite{zhao_vrank_2025}  extend oracle-free repair to functional correctness, leveraging the internal consistency of LLM outputs. By generating multiple candidate implementations and analyzing their behavioral agreement through simulation, they can identify and select functionally correct designs without requiring a golden reference. 

Importantly, both paradigms operate at the implementation level and assume the specification itself is sufficiently correct. In contrast, our work studies the upstream problem of recovering intended functionality when the specification itself contains semantic defects.

To address this challenge, we investigate specification repair in this work with two fundamentally different paradigms for recovering design intent from defective specifications.
The first paradigm performs \emph{semantic repair} to directly analyze and edit the specification text to resolve inconsistencies before HDL generation. The second performs \emph{behavioral recovery} to infer intended functionality from behavioral agreement among multiple generated implementations through simulation.


The key contributions of this paper are summarized as follows:

\begin{itemize}

\item
We present the first systematic study of semantically defective specifications in Verilog HDL generation, covering three common defect categories: vagueness, contradiction, and incompleteness.

\item
Two complementary paradigms for recovering design intent from defective specifications are proposed: Spec-Level Repair, which performs semantic inconsistency mining and targeted specification repair, and Sim-Level Repair, which leverages behavioral consensus across simulated implementations without requiring golden testbenches.

\item
We introduce and open-source two benchmark datasets for specification repair in HDL generation: VerilogEval-Defect for single-module tasks and ComplexVDB-Defect for realistic multi-module designs with detailed engineering specifications.




\item
Experimental results demonstrate that spec-level repair achieved 22.3\% accuracy improvement on contradictory prompts on VerilogEval-Defect, such improvement further increased to 28.7\% with sim-level repair. On ComplexVDB-Defect, we achieved 13.7\% overall improvement by applying sim-level repair. In addition, we uncover a scaling divergence in two paradigms: spec-level repair is concise in single-module tasks but degrades as specification complexity increases, while sim-level repair through simulation remains robust across both simple and complex design settings.

\end{itemize}

The rest of this paper is organized as follows. Section~\ref{sec:background} provides background and motivation. Section~\ref{sec:proposed} details the proposed framework. Section~\ref{sec:dataset} details the datasets we constructed. Experimental results and discussions are presented in Section~\ref{sec:experiments}. Section~\ref{sec:conclusion} concludes the paper.

\section{Background and Motivation}
\label{sec:background}


Hardware design is fundamentally a staged refinement process that progressively transforms natural-language specifications into executable implementations~\cite{xu_kangwei_invited_2026}. The specification captures design intent and governs downstream decisions throughout architecture design, RTL implementation, and physical realization.
When the specification is semantically sound, downstream refinement stages can preserve intended behavior. In contrast, defective specifications can propagate incorrect intent throughout the design flow, ultimately producing functionally incorrect hardware.

In practice, specifications written by engineers often contain imperfections due to careless manual design, miscommunication during project management and collaboration, or insufficient consideration of edge cases. As LLMs accelerate hardware design and lower the implementation burden, they increasingly receive prototype-level specifications written with less rigor than final documentation.
A similar trend is receiving rising attention in software engineering. In the software domain, researchers have begun to directly address the problem of defective prompts. Studies have shown that ambiguous, contradictory, and incomplete task descriptions substantially degrade the performance of code LLMs~\cite{larbi_when_2025}. Frameworks such as ClarifyGPT~\cite{mu_clarifygpt_2024} identify ambiguities and solicit targeted human clarifications before code generation. Later, SpecFix~\cite{jia2025automated} leverages human-written test cases to ground the disambiguation process, using input-output examples as a proxy for intent.

However, these methods still rely on a relatively strong external oracle, such as a direct human answer or a human-written golden testbench.
In hardware design, the challenge of defective specifications is particularly acute, where engineering expertise is expensive and comprehensive golden testbenches are rarely available during early specification stages. We therefore investigate whether defective hardware specifications can be repaired using light-weight human interaction of only confirmation on LLM-generated edits, while avoiding dependence on strong external oracles. In other words, we target \textit{minimum-oracle} scenario, and study how and how efficiently we can fix imperfect specification in verilog design.

\begin{figure}
    \centering
    \includegraphics[width=1\linewidth]{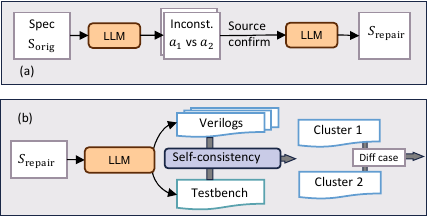}
    \caption{Two repair mechanisms explored in VClare.}
    \label{fig:framework_brief}
\end{figure}

Fig.~\ref{fig:framework_brief} illustrates the two light-weight repair mechanisms explored in VClare, each with a different demand on LLM reasoning.
Spec-Level Repair operates directly on the specification text through inconsistency mining and targeted editing before HDL generation. It requires the LLM's ability to locate inconsistency. Sim-Level Repair performs behavioral recovery after HDL generation by identifying consensus among simulated implementations. It relies on execution-level agreement and LLM's ability to globally implement the specification.
As we later show, Spec-Level Repair is effective on concise specifications but degrades as specifications become more complex, whereas Sim-Level Repair remains robust across both settings.
\section{The Proposed Framework}
\label{sec:proposed}


As illustrated in Fig.~\ref{fig:framework_brief}, VClare combines two complementary mechanisms for recovering design intent from defective specifications:  (a) \textbf{Spec-Level Repair}, which performs semantic inconsistency mining and targeted specification editing prior to RTL generation, and (b) \textbf{Sim-Level Repair}, which performs behavioral recovery through simulation-based clustering and optional lightweight arbitration after RTL generation.

Spec-Level Repair and Sim-Level Repair can operate independently or sequentially. In the sequential configuration, Spec-Level Repair first attempts to restore semantic consistency within the specification, while Sim-Level Repair subsequently performs behavioral validation over generated implementations to recover from residual specification defects or generation errors.


\subsection{Spec-Level Repair: Semantic Inconsistency Mining and Targeted Repair}

Spec-Level Repair targets the specification directly, before any Verilog code is generated. The goal is to detect and resolve semantic defects at the source with minimal human intervention. The process consists of three steps: inconsistency mining, human confirmation, and targeted repair. An example of spec-level repair is presented in Appendix I.

\textit{Semantic inconsistency mining.} Given an original specification $S_{\text{orig}}$, the LLM is prompted to systematically analyze the text for semantic defects. For each potential defect, the model extracts an \textit{inconsistency pair} $(a_1, a_2)$, where $a_1$ and $a_2$ are two statements in the specification that appear to conflict, or that together reveal a vagueness or incompleteness. The LLM is required to identify up to three such pairs per specification. If a vagueness or incompleteness is detected but no second source can be located to form a pair, the model sets $a_1 = a_2$, marking the defect as a standalone issue that still requires human attention.

Formally, the output of this step is a set of pairs $\mathcal{P} = \{(a_1^{(i)}, a_2^{(i)})\}_{i=1}^{m}$, where $m \leq 3$, and each pair captures a suspected inconsistency or underspecification in $S_{\text{orig}}$. All $m$ inconsistency pairs are extracted in a single conversation with LLM, ensuring that they share the same context and thus avoid repetition to the best of LLM's effort. The detailed m value discussion with different defect types and data complexity is available in Appendix III.

The extracted pairs $\mathcal{P}$ are presented to a human engineer. Crucially, the human is \textit{not} asked to provide corrections, write test cases, or supply missing specifications. The only response required is a simple confirmation that there is a source it should trust. \textit{Source 1 is correct:} $a_1$ reflects the intended behavior; $a_2$ should be modified. \textit{Source 2 is correct:} $a_2$ reflects the intended behavior; $a_1$ should be modified. \textit{Irrelevant:} The pair does not represent a genuine defect and should be discarded, or the pair has no correction info for that defect.

\subsubsection{Targeted Repair}

For each validated inconsistency pair, the LLM produces a constrained edit that modifies only the incorrect statement while preserving unrelated specification content unchanged.
Pairs marked as irrelevant are discarded without modification. The result is a repaired specification, denoted $S_{\text{repair}}$.

\subsection{Sim-Level Repair: Behavior Clustering with Optional Confirmation}

Unlike Spec-Level Repair, Sim-Level Repair does not attempt to modify or reinterpret the specification text itself. Instead, it performs intent recovery behaviorally by identifying consensus among multiple generated implementations through simulation. It adopts the VRank pipeline~\cite{zhao_vrank_2025}—execution-based clustering, MBR-based ranking, and consensus selection—and augments it with an optional arbitration step that activates when multiple viable clusters emerge and a human is available to provide a minimal confirmation signal.

\subsubsection{Code Candidate Generation and Cluster ranking}

The framework generates $N$ Verilog code candidates $\mathcal{C} = \{c_1, c_2, \ldots, c_N\}$ by repeatedly prompting LLM with temperature above 0 from the specification $S$ (where $S = S_{\text{repair}}$ if Spec-Level Repair is applied, or $S = S_{\text{orig}}$ otherwise). An automated testbench $T$ containing multiple test cases is also generated via LLM, with extra prompting to cover extra cases because the prompt may be defective. Each candidate $c_i$ is simulated on $T$ using Icarus Verilog. Candidates that fail to compile or produce no output are each placed in singleton clusters.

Candidates that successfully simulated are grouped by behavioral equivalence: $c_i$ and $c_j$ belong to the same cluster if and only if their outputs match on all test cases. Let the resulting clusters be $\{C_1, C_2, \ldots, C_k\}$.

Clusters are ranked using the Minimum Bayes Risk (MBR)~\cite{kumar_minimum_2004} consistency score, as defined in VRank~\cite{zhao_vrank_2025}. The score of a candidate $c$ is:
$R(c) = n - \sum_{c' \in \mathcal{C}} \ell(c, c')$,
where $\ell(c, c') = 1$ if $c$ and $c'$ differ on any test case (or if either fails simulation), and $0$ otherwise. Clusters are ranked by their scores in descending order, with $C_1$ denoting the top-ranked cluster.

\subsubsection{Optional Confirmation on Distinguishing Test Cases}

When multiple clusters contain successfully simulated designs (i.e., $k > 1$), the framework identifies the first test case $t_d \in T$ on which $C_1$ and $C_2$ produce divergent outputs. At this point, an optional human confirmation step is available. Human engineers can step in and confirm on the desired behavior under this test case. If no human steps in, the Sim-level repair degrades to standard VRank selection without arbitration.

\subsection{Complementarity of the Two Paradigms}

The two intent-recovery paradigms address defective specifications from complementary perspectives.

Spec-Level Repair attempts to restore semantic consistency directly at the specification layer, which is highly effective when defects are localized, and inconsistency localization remains reliable. However, this paradigm depends heavily on accurate long-context reasoning over natural-language specifications.

Sim-Level Repair instead performs behavioral recovery through execution-level consensus, avoiding direct modification of the specification itself. As a result, it is substantially more robust to specification length and structural complexity.

In practice, these paradigms can be combined sequentially for concise specifications where semantic inconsistency localization remains reliable. However, as the complexity of the specification increases, our experiments show that applying behavioral recovery directly to the original specification is often the most robust strategy.
\section{New Datasets for Testing Imperfect Specifications of Circuit Design}
\label{sec:dataset}

To provide a systematic evaluation of the specification defect problem in Verilog generation, we study the following three defects that commonly appear in requirement engineering~\cite{montgomery_empirical_2022, larbi_when_2025}, by injecting them into correct specifications.

\begin{itemize}
\item \textit{Contradiction:} There are conflicting statements in different sections of the specification (e.g., specifying both synchronous and asynchronous reset behavior, or conflicting signal polarity requirements).
\item \textit{Incompleteness:} Descriptions of some edge-case behavior or constraints are missing. (e.g., overflow handling, reset behavior, or invalid input processing) from otherwise complete specifications.
\item \textit{Vagueness:} Precise behavioral descriptions are absent, the specification uses underspecified alternative descriptions or uses words that have multiple meanings.

\end{itemize}

\begin{figure}[t]
    \centering
    \includegraphics[width=0.85\linewidth]{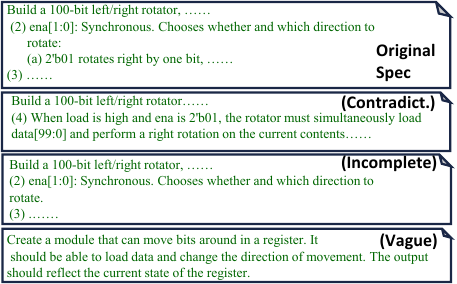}
    \caption{Specification example of original prompts and three defects}
    \label{fig:defects_example}
\end{figure}

Fig.~\ref{fig:main_defect_injection} shows an example with the original prompts and the prompts after defect injection. Detailed examples of defect injection requirement prompts and general prompts are presented in Appendix II. We manually check each generated specification and make modifications if it is unrealistic.

By systematically injecting these defects into human-crafted functional specifications, we construct and open-source two benchmark datasets designed to assess specification repair across different levels of design complexity. 

\textbf{VerilogEval-Defect (Single-Module):} Derived from the VerilogEval-human benchmark~\cite{liu_invited_2023}, which comprises 156 single-module Verilog generation tasks. For each task, we systematically inject semantic defects of three types — vagueness, contradiction, and incompleteness — into the original human-written specification to create a defective counterpart. An example of our defect injection script is presented in \ref{app:2_defect_inject}. This controlled setup enables fine-grained evaluation of how each defect type impacts generation quality and how effectively our repair methods recover correct functionality. For each task, the original unmodified specification serves as a reference, and a golden testbench from the VerilogEval benchmark is used exclusively for final pass/fail evaluation (not during the repair process).

\begin{table}[t]
\vspace{-4pt}
\centering
\footnotesize
\caption{Design domains of ComplexVDB-Defect.}
\vspace{-6pt}
\label{tab:ComplexVDB-domains}
\begin{tabular}{@{}l| c| >{\raggedright\arraybackslash}p{0.45\columnwidth}@{}}
\toprule
\textbf{Design Domain} & \textbf{Count} & \textbf{Examples} \\
\midrule
Arithmetic \& Datapath         & 8x3  & ALU, Wallace\_multiplier \\
Bus \& Interconnect            & 9x3  & AHB\_master, AXI\_lite\_slave \\
Cryptography \& Security       & 4x3  & AES128, CRC\_16\_usb \\
Memory \& Buffers              & 5x3  & SDRAM, FIFO, LIFObuffer \\
Peripherals \& Input/Output    & 10x3 & UART\_tx, SPI, DAC \\
Processor \& CPU               & 9x3  & CtrlUnit, E203\_reset\_ctrl \\
Signal Processing \& Accelerator & 5x3 & CIC, FFT, Filt\_cicd \\
Storage \& SD Card             & 3x3  & SD\_bd, SD\_rx\_fifo \\

\bottomrule
\end{tabular}
\vspace{-10pt}
\end{table}

\textbf{ComplexVDB-Defect (Multi-Module):} To evaluate specification repair under more realistic engineering conditions, we introduce ComplexVDB-Defect. The original specifications of ComplexVDB are collected from~\cite{zuo_complexvcoder_2025}, and ~\cite{junzhe_liu_reflectbench_2026}. As detailed in Table~\ref{tab:ComplexVDB-domains}, ComplexVDB-Defect consists of 53x3 real-world multi-module Verilog generation tasks across 8 distinct domains. Each task includes a detailed specification with explicit port definitions and behavioral descriptions, reflecting the complexity and verbosity of real-world hardware design documents. For these tasks, the specifications are sufficiently long and detailed that LLM-based global reasoning becomes challenging. As with VerilogEval-Defect, we inject semantic defects into these specifications for controlled evaluation. Golden testbenches are used only for final evaluation.

\begin{figure}[t]
    \centering
    \includegraphics[width=1\linewidth]{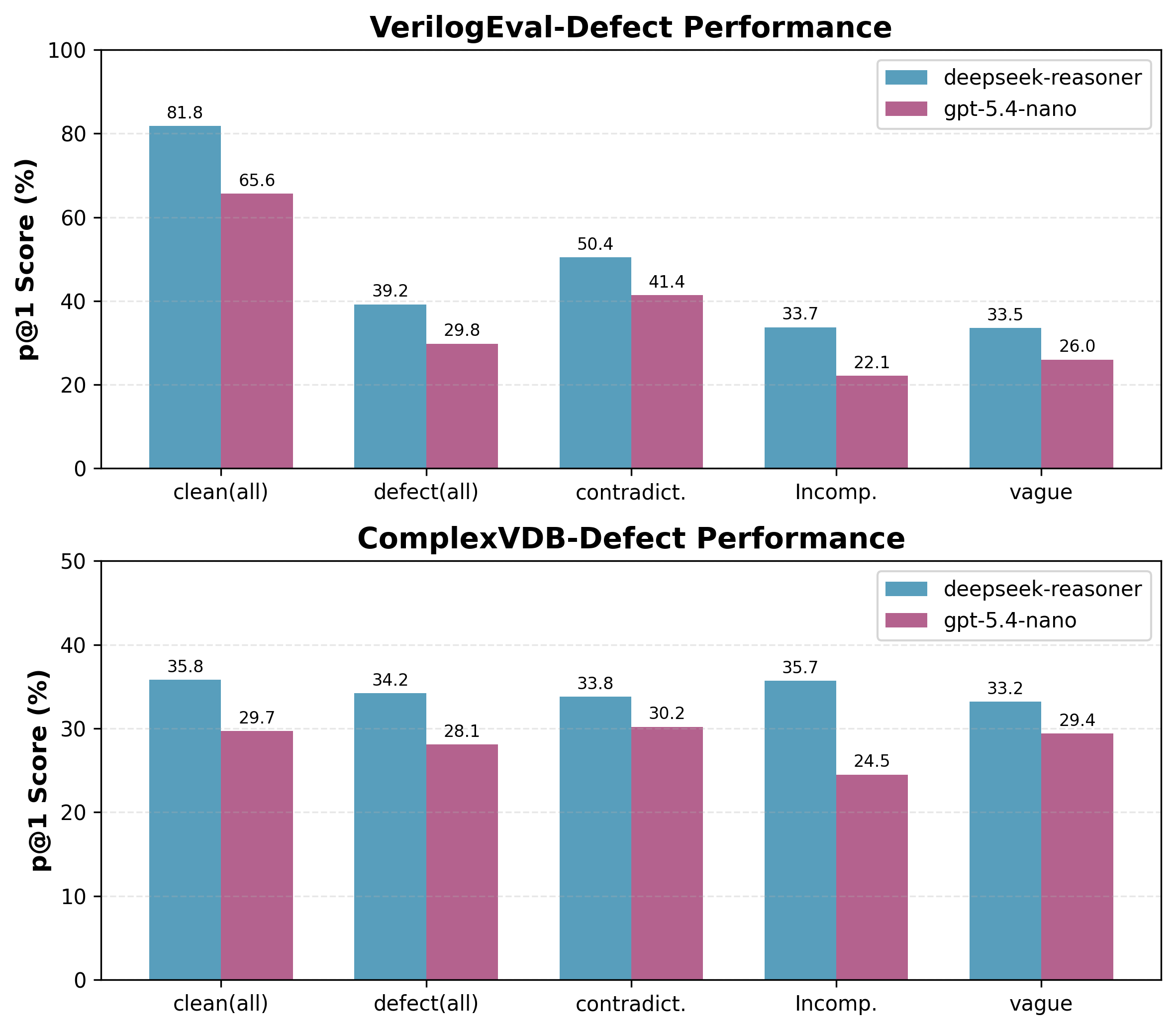}
    \caption{Dataset accuracy after Defect Injection}
    \label{fig:dataset_accuracy_loss}
\end{figure}

After the injection, the two datasets both showed accuracy loss on the code generated on the imperfect verilog module specification. In Fig.~\ref{fig:dataset_accuracy_loss}, we report pass@1, the possibility that one generated code successfully passed the original testbench.


\section{Experimental Evaluation}
\label{sec:experiments}


\subsection{Models and Implementation}

All experiments are conducted using deepseek-v4-flash~\cite{noauthor_deepseek-v4_nodate} and OpenAI gpt-5.4-nano~\cite{noauthor_introducing_2026} as the backbone LLM for both specification repair and Verilog generation. In the remainder of the section, we use DS and GPT as abbreviations. We use the default temperature, and the reasoning effort for deepseek-v4-flash and gpt-5.4-nano is set to low and medium, respectively. Simulation is performed using Icarus Verilog (iverilog)~\cite{williams_steveicarusiverilog_2024} v13.0, on a server with 2 Xeon Gold 6126 CPUs and 280 GB RAM. Each experimental configuration is run with 5 runs to account for LLM non-determinism.

\subsection{Evaluation Metrics}

Our primary evaluation metric is pass@k, as defined in~\cite{chen2021evaluating}, in which a pass is recorded if any of the top-k generated implementations passes the golden testbench. In random sampling, Pass@k is given by the formulation below:

\begin{equation}
\text{pass@k} := \mathbb{E}_{\text{Problems}} \left[ 1 - \frac{\binom{n-c}{k}}{\binom{n}{k}} \right]
\end{equation}
where \(n\) represents the number of sampled candidates, and \(c\) is the count of correct ones among them. For a fair comparison, we set $n=10$ in all experiments. After clustering, the pass@k is indicated by whether or not one of the picked k samples contains a code that passes the golden testbench.

\subsection{Baselines and Evaluated Repair Configurations}

We compare VClare against the following baselines.

\textit{No Repair (Original):} Verilog is generated directly from the defective specification without any repair or correction process.

\textit{Blind Fix:} The LLM is informed that the specification may contain defects and is prompted to repair the specification in a zero-shot manner before Verilog generation. No structured inconsistency mining or behavioral validation is provided. Pass@1 is estimated with $n=10$ generations.

\textit{VRank:} Behavioral clustering using the original VRank~\cite{zhao_vrank_2025} pipeline. Multiple Verilog candidates are generated from the defective specification and ranked through execution-based behavioral consensus without test-case arbitration.

\textit{SpecFix-no-oracle:} An adaptation of the software-domain specification repair framework SpecFix~\cite{jia2025automated} to Verilog generation. The method first generates multiple Verilog implementations from the defective specification, selects the dominant behavioral cluster, revises the specification based on the selected implementations, and finally regenerates Verilog from the revised specification. This baseline represents an implementation-first specification repair strategy.

In addition to these baselines, we evaluate the following repair configurations within the VClare framework.

\textit{Spec-Level Repair:} The specification-level inconsistency mining and targeted specification repair pipeline described in Section~\ref{sec:proposed} is applied prior to Verilog generation. A single Verilog implementation is then generated from the repaired specification. Pass@1 is estimated with $n=10$ generations.

\textit{Sim-Level Repair:} The simulation-based behavioral clustering pipeline described in Section~\ref{sec:proposed} is applied directly on Verilog candidates generated from the original defective specification. Candidate implementations are ranked through behavioral consensus, with optional lightweight test-case arbitration.

\textit{Hybrid Repair (NA):} Spec-Level Repair is first applied to produce a repaired specification, followed by Sim-Level Repair using standard VRank behavioral selection without test-case arbitration.

\textit{Hybrid Repair:} The full sequential configuration combining Spec and Sim level repair, with lightweight test-case arbitration as described in Section~\ref{sec:proposed}.
\subsection{Results and Analysis}


\begin{table}[t]
\centering
\caption{Results on VerilogEval-Defect (accuracy \% in pass@1). Best results in bold.}
\label{tab:single_brief}
\small
\begin{tabular}{lcccccccc}
\toprule
\multirow{2}{*}{\textbf{Method}} & \multicolumn{2}{c}{\textbf{Overall}} & \multicolumn{2}{c}{\textbf{Incomp.}} & \multicolumn{2}{c}{\textbf{Vague}} & \multicolumn{2}{c}{\textbf{Contradict.}} \\
\cmidrule(lr){2-3} \cmidrule(lr){4-5} \cmidrule(lr){6-7} \cmidrule(lr){8-9}
& \textbf{DS} & \textbf{GPT} & \textbf{DS} & \textbf{GPT} & \textbf{DS} & \textbf{GPT} & \textbf{DS} & \textbf{GPT} \\
\midrule
No Repair & 39.2 & 29.8 & 50.4 & 41.4 & 33.7 & 22.1 & 33.5 & 26.0 \\
Blind Fix & 20.8 & 26.0 & 20.3 & 33.5 & 12.4 & 14.8 & 29.8 & 29.7 \\
VRank & 41.6 & 31.7 & 54.2 & 44.6 & 34.2 & 23.0 & 36.3 & 27.6 \\
\makecell{SpecFix\\ (no-oracle)} & 41.4& 31.0& 54.7& 41.7& 34.6& 25.0& 35.1& 26.2\\
Spec-level & 44.0 & 36.0 & 44.7& 38.1 & 31.8& 21.4 & 55.6& 48.5 \\
Sim-level & 47.7 & 38.4 & \textbf{58.1} & \textbf{48.8} & \textbf{41.3} & \textbf{29.0} & 43.7 & 37.4 \\
\makecell{Hybrid\\ (NA)} & 45.8& 38.8 & 46.9& 41.9 & 32.7& 22.4 & 57.8& 52.1 \\
Hybrid & \textbf{50.0}& \textbf{44.4} & 52.7& 47.4 & 37.9& 28.5 & \textbf{59.5}& \textbf{57.4} \\
\bottomrule
\end{tabular}
\vspace{-6pt}
\end{table}

Table~\ref{tab:single_brief} presents the results of the VerilogEval-Defect dataset.

\textbf{Finding 1: Specification-level repair is effective for contradiction defects.} As shown in Table~\ref{tab:single_brief}, Spec-level Repair showed mixed results across defect types. On contradiction defects, it delivered substantial gains: +22.1\% on DeepSeek and +22.5\% on GPT over No Repair, indicating that the LLM can reliably identify conflicting statements and resolve them when the specification is concise. However, on incompleteness defects and Vague defects, Spec-level Repair underperformed No Repair. We attribute this to the nature of incompleteness and vagueness: the fixes are harder because LLM lacks information to correct. Therefore, it is a better idea to step in when vagueness or incompleteness is found. We further discuss this phenomenon in Appendix III. Blind Fix consistently performed worst across all defect types, confirming that unstructured, non-localized repair without inconsistency mining is ineffective.

\textbf{Finding 2: Sim-Level Repair provides consistent improvements across all defect types.} In contrast to Spec-level Repair's uneven performance, Sim-level Repair improved over No Repair in every defect category, with particularly strong gains in incompleteness (+7.7\% DS, +7.4\% GPT) and vagueness (+7.6\% DS, +6.9\% GPT). This confirms that behavioral clustering can recover functional correctness even when the LLM cannot explicitly identify what is wrong with the specification. The software domain SoTA work SpecFix-no-oracle baseline consistently underperformed Sim-level, showing that the additional spec fix after clustering provided an extra error propagation level.

\textbf{Finding 3: Hybrid repair achieves the best overall result, with Spec-level and Sim-level contributing complementarily.} Hybrid repair achieved the highest overall Pass@1 on both models (50.0\% DS, 44.4\% GPT). Meanwhile, where Spec-level excelled, the full framework largely preserved those gains (59.5\% DS, 57.4\% GPT). On incompleteness, the Sim-Level component partially compensated for Spec-Level's weakness (52.7\% DS, 47.4\% GPT, vs. Sim-level's 58.1\% and 48.8\%).


\begin{table}[t]
\centering
\vspace{-8pt}
\caption{Results on ComplexVDB-Defect (accuracy \% in pass@1). Best results in bold.}
\label{tab:multi_brief}
\small
\begin{tabular}{lcccccccc}
\toprule
\multirow{2}{*}{\textbf{Method}} & \multicolumn{2}{c}{\textbf{Overall}} & \multicolumn{2}{c}{\textbf{Incomp.}} & \multicolumn{2}{c}{\textbf{Vague}} & \multicolumn{2}{c}{\textbf{Contradict.}} \\
\cmidrule(lr){2-3} \cmidrule(lr){4-5} \cmidrule(lr){6-7} \cmidrule(lr){8-9}
& \textbf{DS} & \textbf{GPT} & \textbf{DS} & \textbf{GPT} & \textbf{DS} & \textbf{GPT} & \textbf{DS} & \textbf{GPT} \\
\midrule
No Repair & 34.2 & 28.1 & 33.8 & 30.2 & 35.7 & 24.5 & 33.2 & 29.4 \\
Blind Fix & 13.0 & 17.5 & 11.5 & 19.8 & 14.5 & 14.7 & 12.8 & 18.1 \\
VRank & 42.4 & 35.0 & 43.2 & 36.5 & 46.3 & 32.8 & 37.7 & 35.8 \\
\makecell{SpecFix\\ (no-oracle)} & 28.7& 25.1& 33.8& 28.1& 28.2& 26.3& 23.9& 20.8\\
Spec-level & 23.6 & 24.0 & 22.8 & 26.0 & 26.4 & 20.4 & 21.7 & 25.5 \\
Sim-level & \textbf{49.8} & \textbf{39.9} & \textbf{50.5} & \textbf{42.8} & \textbf{49.8} &\textbf{ 32.8 }& \textbf{49.3} & \textbf{44.1} \\
\makecell{Hybrid\\ (NA)} & 30.8 & 30.2 & 32.7 & 31.1 & 30.7 & 25.0 & 29.0 & 34.4 \\
Hybrid & 35.4 & 32.9 & 37.6 & 33.2 & 35.6 & 27.8 & 33.1 & 37.7 \\
\bottomrule
\end{tabular}
\vspace{-6pt}
\end{table}
Table~\ref{tab:multi_brief} presents the results on the ComplexVDB-Defect dataset.

\begin{figure}
    \centering
    \vspace{-12pt}
    \includegraphics[width=0.9\linewidth, trim=10 5 10 5, clip]{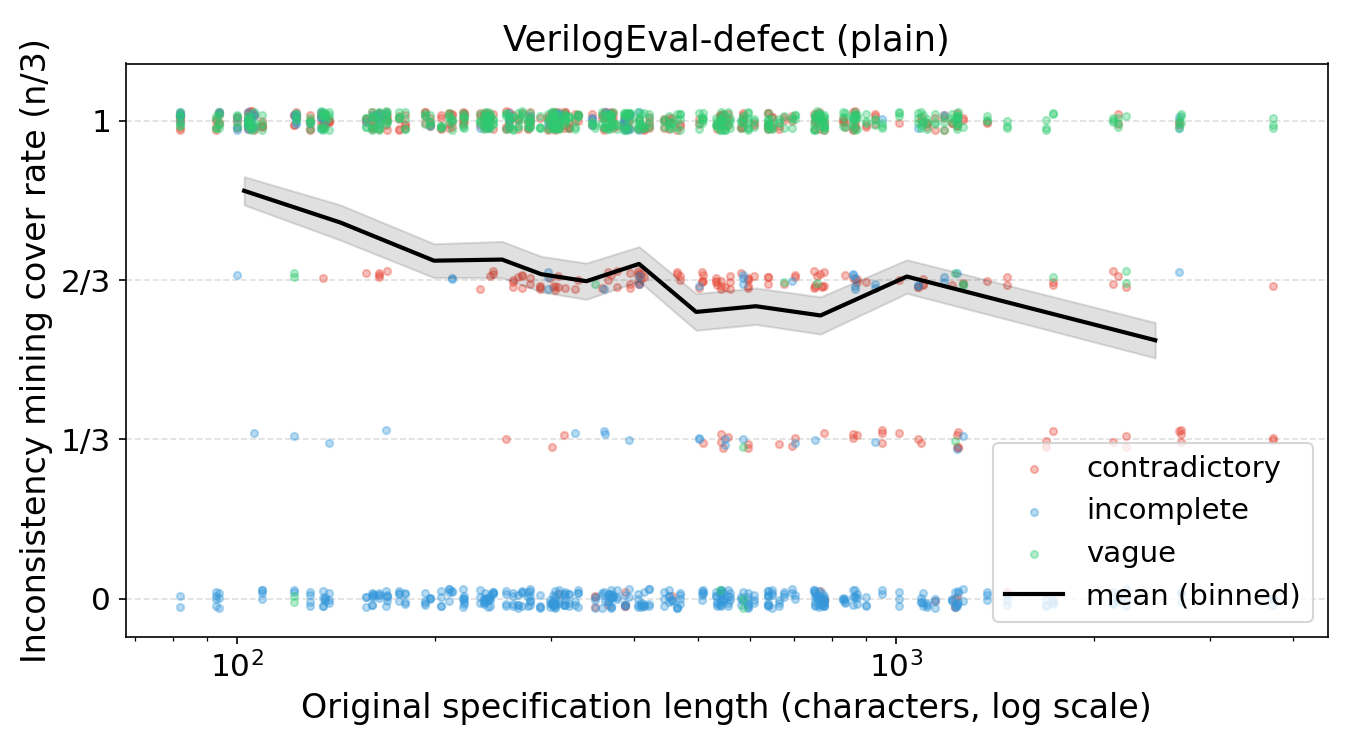}
    \caption{Inconsistency mining cover rate on VerilogEval-Defect, by specification length.}
    \label{fig:cover_rate_single}
    \vspace{-3pt}
\end{figure}

\textbf{Finding 4: Specification-level repair becomes counterproductive as design complexity increases.} On ComplexVDB-Defect, Spec-level Repair reduced Pass@1 compared to No Repair across all defect categories (Overall: 23.6\% vs. 34.2\% on DS, 24.0\% vs. 28.1\% on GPT). Blind Fix degraded even further (13.0\% DS, 17.5\% GPT), confirming the danger of unguided repair on long specifications. We identify two contributing factors for the Spec-level's performance degradation.
First, inconsistency localization becomes increasingly difficult as specifications grow longer. We define the cover rate of inconsistency mining as the fraction of mined inconsistency pairs that cover the injected defect. As shown in Fig.~\ref{fig:cover_rate_single}, the cover rate declines steadily with specification length on VerilogEval-Defect.
Second, complex multi-module specifications contain substantial structural and behavioral redundancy. Compared to single-module prompts, ComplexVDB-Defect specifications include detailed connectivity descriptions, interface definitions, sub-module interactions, and repeated behavioral constraints. As a result, many injected defects become partially compensatable by the surrounding context. Consequently, defects become harder to distinguish from normal specification redundancy.

\begin{figure}
    \centering
    \vspace{-12pt}
    \includegraphics[width=0.9\linewidth, trim=10 5 10 5, clip]{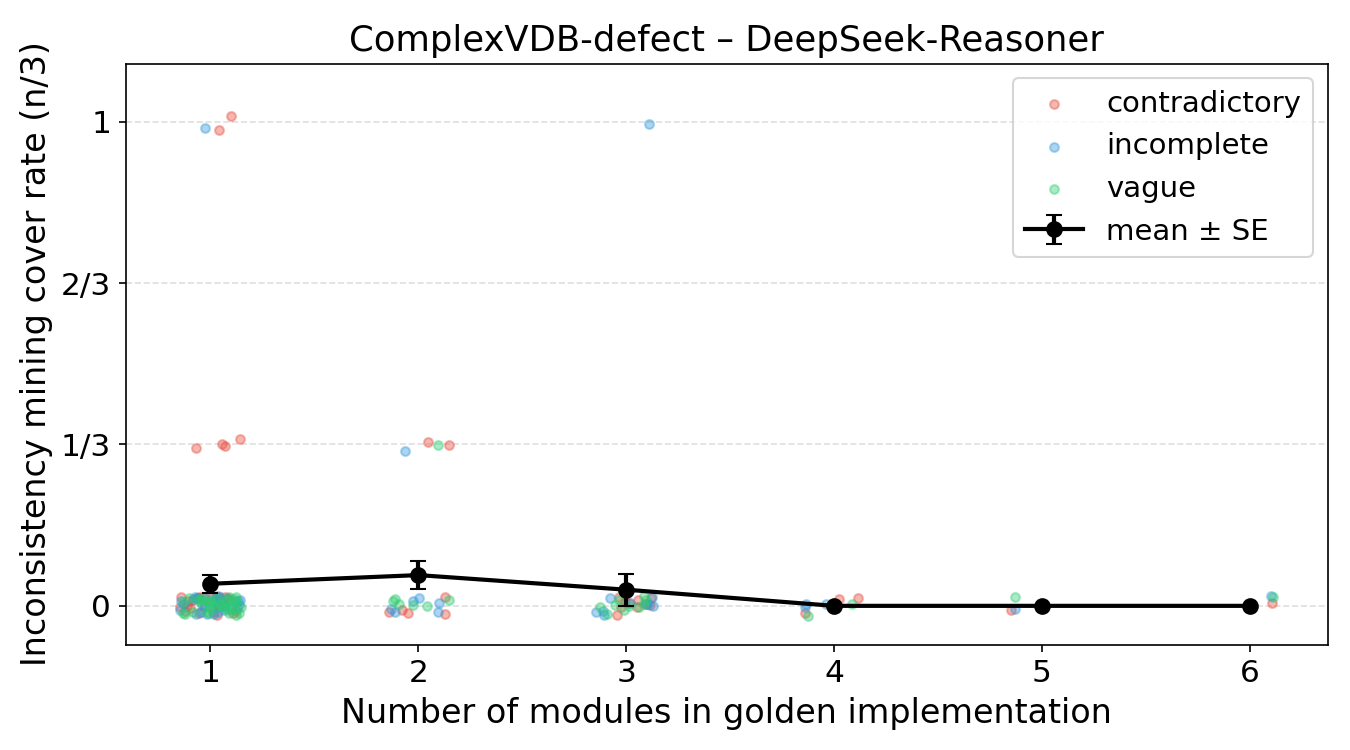}
    \caption{Inconsistency mining cover rate on ComplexVDB-Defect, by number of sub-modules.}
    \label{fig:cover_rate_by_module}
    \vspace{-16pt}
\end{figure}

This phenomenon is reflected in two observations. First, as shown in Fig.~\ref{fig:dataset_accuracy_loss}, the accuracy degradation caused by defect injection was substantially smaller on ComplexVDB-Defect (35.8\% drop to 34.2\% for DS pass@1) than on VerilogEval-Defect(81.8\% drop to 39.2\% for DS pass@1), indicating that complex specifications with more redundancy are inherently more fault-tolerant. Second, as shown in Fig.~\ref{fig:cover_rate_by_module}, inconsistency mining accuracy further declined as the number of sub-modules increased. The richer the surrounding context becomes, the more difficult it is for the LLM to isolate which statement is genuinely defective rather than merely underspecified or implicitly defined elsewhere.

\textbf{Finding 5: Sim-Level Repair remains robust and dominates on complex specifications.} In stark contrast to Spec-level Repair, Sim-level Repair achieved the highest Pass@1 across all defect categories on ComplexVDB-Defect (Overall: 49.8\% DS, 39.9\% GPT), outperforming No Repair by 15.6\% and 11.8\% respectively. VRank alone also improved substantially over No Repair (+8.2\% DS, +6.9\% GPT), confirming that behavioral clustering is inherently robust to specification complexity. Because these paradigms do not require the LLM to precisely locate and edit defects, they avoid the pitfalls that derail Spec-Level Repair.

\textbf{Finding 6: Spec-Level Repair can harm downstream Sim-Level performance on complex tasks.} VClare (Full), which applied Spec-Level Repair before Sim-Level Repair, achieved 35.4\% (DS) and 32.9\% (GPT) overall—a modest improvement over No Repair, but substantially below Sim-level Repair. This indicates that when Spec-Level Repair introduces spurious modifications to an already-adequate specification, the downstream Sim-Level Repair can only partially recover. VClare (NA) performed similarly (30.8\% DS, 30.2\% GPT), further from Sim-level. The SpecFix-no-oracle baseline degrades below No Repair (28.7\% DS, 25.1\% GPT), confirming that implementation-first repair strategies are particularly brittle on complex hardware specifications. These results underscore a practical guidance: on complex multi-module designs, Sim-level Repair without prior specification modification is the more reliable configuration.

\subsection{Discussion}
Our results reveal a clear difference between the two repair paradigms' effectiveness as a function of specification complexity. Spec-Level Repair is a solution based on locality. It excels when specifications are concise, defects are localized, and specifications contain explicit contradictions, where the LLM can reliably identify and resolve conflicting statements. However, as specification complexity grows, the detectability of defects deteriorates, and the risk of spurious edits outweighs the benefits of repair. Sim-Level Repair, grounded in behavioral consensus rather than specification comprehension on bug localization, remains robust across the complexity spectrum.

These findings suggest a practical best practice: decomposing large specifications into smaller, self-contained sub-module descriptions. Smaller specifications are not only easier for LLMs to generate from, but also easier to repair—defects are more readily isolated, and the risk of spurious edits is reduced. When decomposition is impractical, applying Sim-Level Repair directly on the original specification, without prior Spec-Level Repair, is the safer and more effective strategy. More broadly, our results establish behavioral validation as an essential safeguard: even when specification-level reasoning fails, behavioral consensus can recover the intended functionality without requiring the LLM to fully locate the defects of the specification.
\section{Conclusion}
\label{sec:conclusion}

This paper presented the first systematic study of semantically defective specifications in Verilog RTL generation. We categorized specification defects into vagueness, contradiction, and incompleteness, and proposed VClare, a framework integrating two complementary repair paradigms: Spec-Level Repair, which performs LLM-driven inconsistency mining directly on the specification text, and Sim-Level Repair, which leverages simulation-based behavioral clustering to recover functional correctness without requiring golden references. Experiments on two newly constructed benchmarks revealed a nuanced finding: Spec-Level Repair is effective for single-module tasks with contradiction defects, but becomes counterproductive on complex multi-module designs, where LLMs are overwhelmed by specification detail. In contrast, Sim-Level Repair remains robust across all complexity levels, establishing behavioral validation as a critical safeguard against LLM reasoning failures in specification repair. Our datasets are publicly released to facilitate future research on this underexplored problem.

\clearpage
\bibliographystyle{ACM-Reference-Format}
\bibliography{VClare}
\balance

\clearpage
\setcounter{section}{0}  
\renewcommand{\thesection}{\Alph{section}}  
\section*{Appendix I: An example of the proposed LLM-driven prompt repair framework for Verilog}
\label{app:1_prompt_repair}

The following is an example extracted from our VerilogEval-defect dataset, Prob105\_rotate100. Due to the length, we only show how incomplete is fixed by gpt-5.4-nano here.

\setlength{\parindent}{0pt}

\begin{lstlisting}[ basicstyle=\ttfamily\footnotesize, breaklines=true, frame=single]
===== ORIGINAL PROMPT =====

Build a 100-bit left/right rotator, with synchronous load and left/right enable. A rotator shifts-in the shifted-out bit from the other end of the register, unlike a shifter that discards the shifted-out bit and shifts in a zero. If enabled, a rotator rotates the bits around and does not modify/discard them.

  (1) load: Loads shift register with data[99:0] instead of rotating. Synchronous active high.
  (2) ena[1:0]: Synchronous. Chooses whether and which direction to rotate:
      (a) 2'b01 rotates right by one bit,
      (b) 2'b10 rotates left by one bit,
      (c) 2'b00 and 2'b11 do not rotate.
  (3) q: The contents of the rotator.
\end{lstlisting}

The injected default add a contradictory line in the prompt, making it:
\begin{lstlisting}[ basicstyle=\ttfamily\footnotesize, breaklines=true, frame=single]
Build a 100-bit left/right rotator, with synchronous load and left/right enable...<same omitted>...

(4) When load is high and ena is 2'b01, the rotator must simultaneously load data[99:0] and perform a right rotation on the current contents, combining both operations in a single clock cycle.
\end{lstlisting}
This prompt has a pass@1 of only 10\%.

Then on inconsistency mining stage, three pairs are proposed by LLM:
\begin{lstlisting}[ basicstyle=\ttfamily\footnotesize, breaklines=true, frame=single]
Pair 1: irrelevant - LLM found a non-issue between header and ena, and rejected by engineer's arbitration.

Pair 2: 
source1: (1) load: Loads shift register with data[99:0] instead of rotating. Synchronous active high.
source2: (4) When load is high and ena is 2'b01, the rotator must simultaneously load data[99:0] AND perform a right rotation - combining both operations in a single clock cycle.
\end{lstlisting}
This is relevant, and engineer point to (source 1). This information is passed back to LLM, and it correctly resolves conflict load vs. simultaneous-rotate conflict in favour of source 1 (load only). 
Since Pair 2 already found a relevant pair of inconsistency, Pair 3 is not checked by engineer.

After the fix, clause (4) is successfully removed by LLM and the fixed prompt. All 10 verilogs generated by this prompt passes the golden testbench.

\section*{Appendix II: Examples of dataset construction prompts and the dataset}
\label{app:2_defect_inject}

\begin{figure}[h]
    \centering
    \includegraphics[width=1\linewidth]{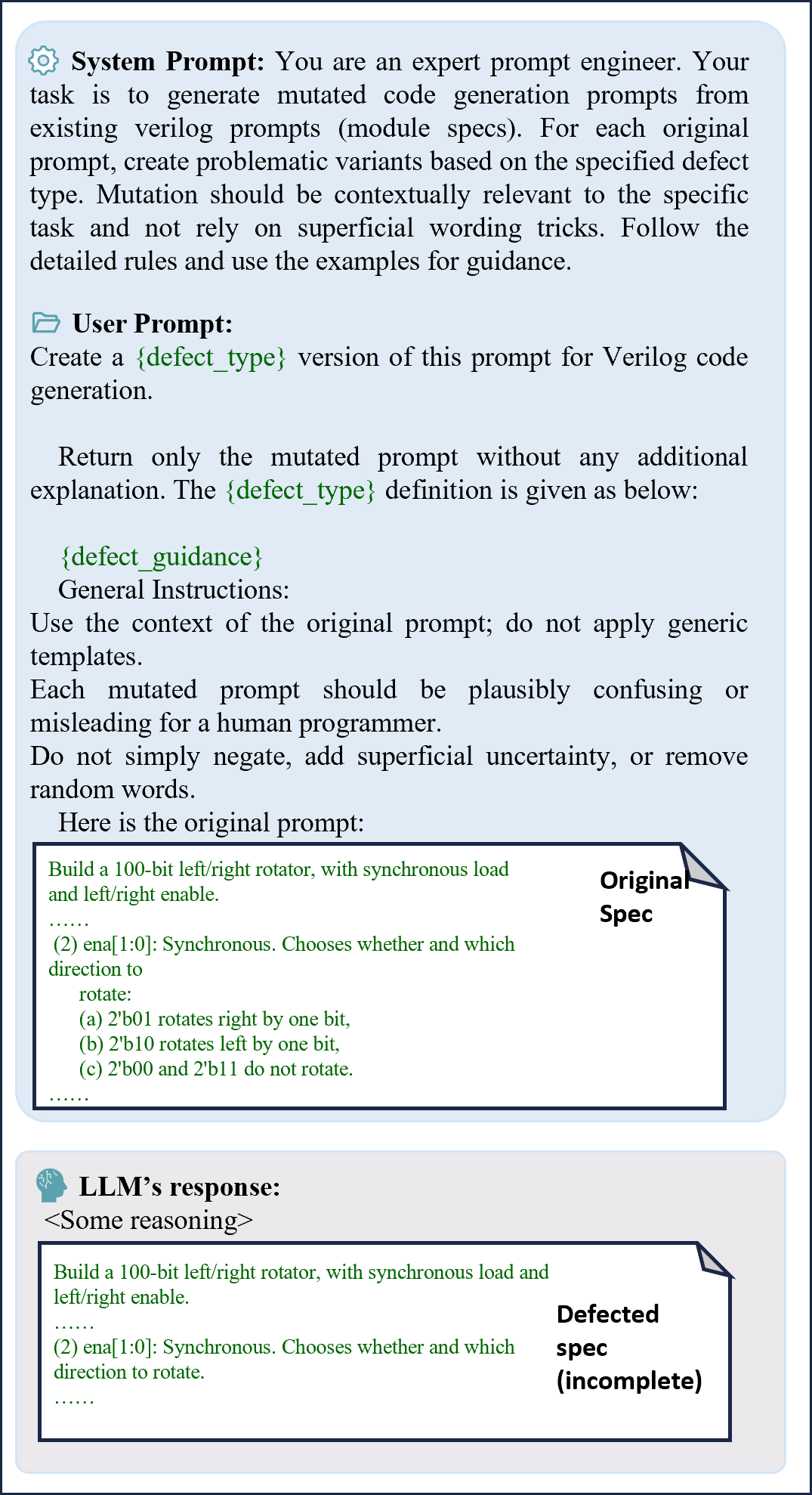}
    \caption{Main defect injection prompt}
    \label{fig:main_defect_injection}
\end{figure}

\begin{figure}[ht]
    \centering
    \includegraphics[width=1\linewidth]{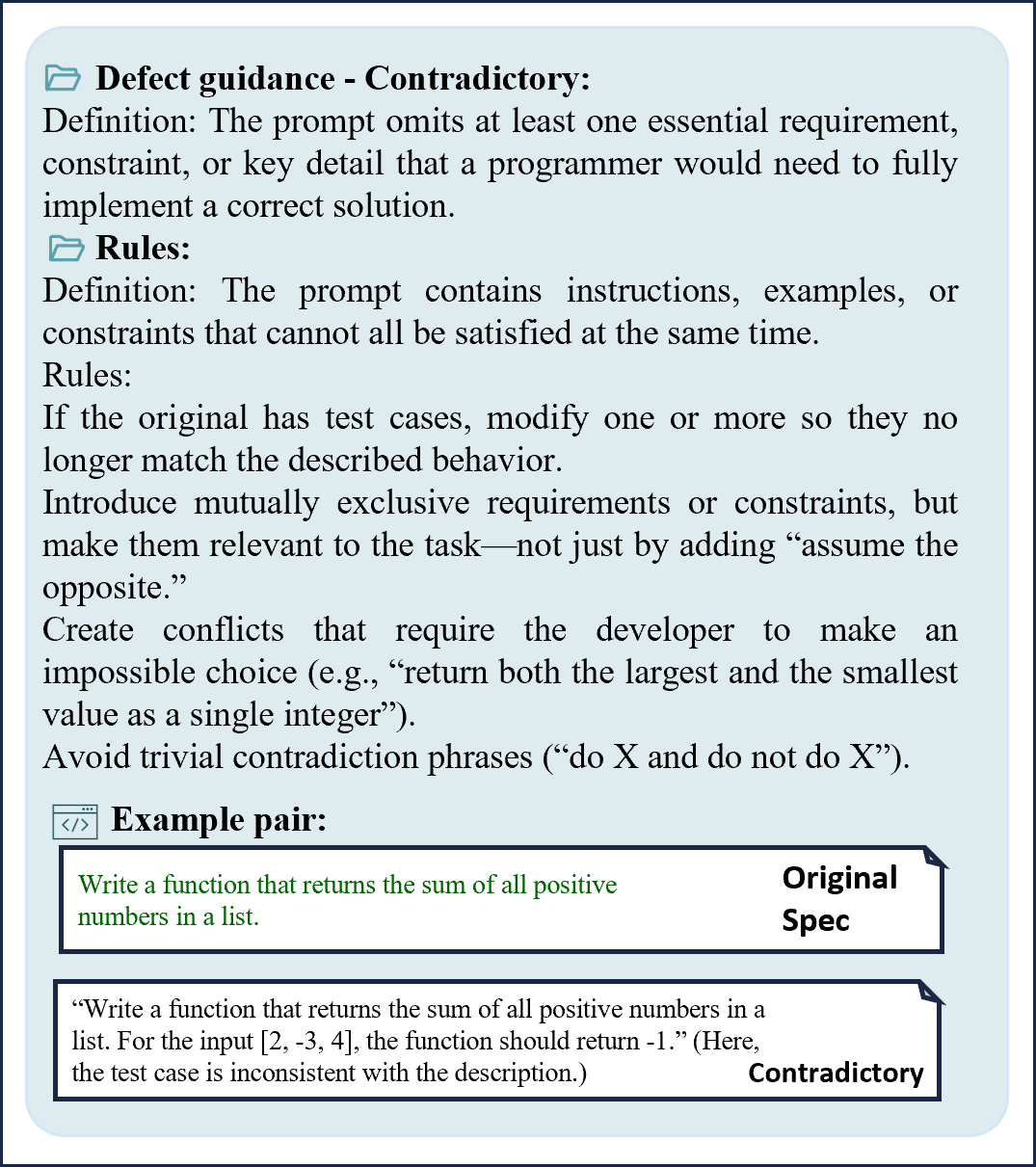}
    \caption{Prompt for contradiction defect injection.}
    \label{fig:defect_injection_contradictory}
\end{figure}

\begin{figure}[ht]
    \centering
    \includegraphics[width=1\linewidth]{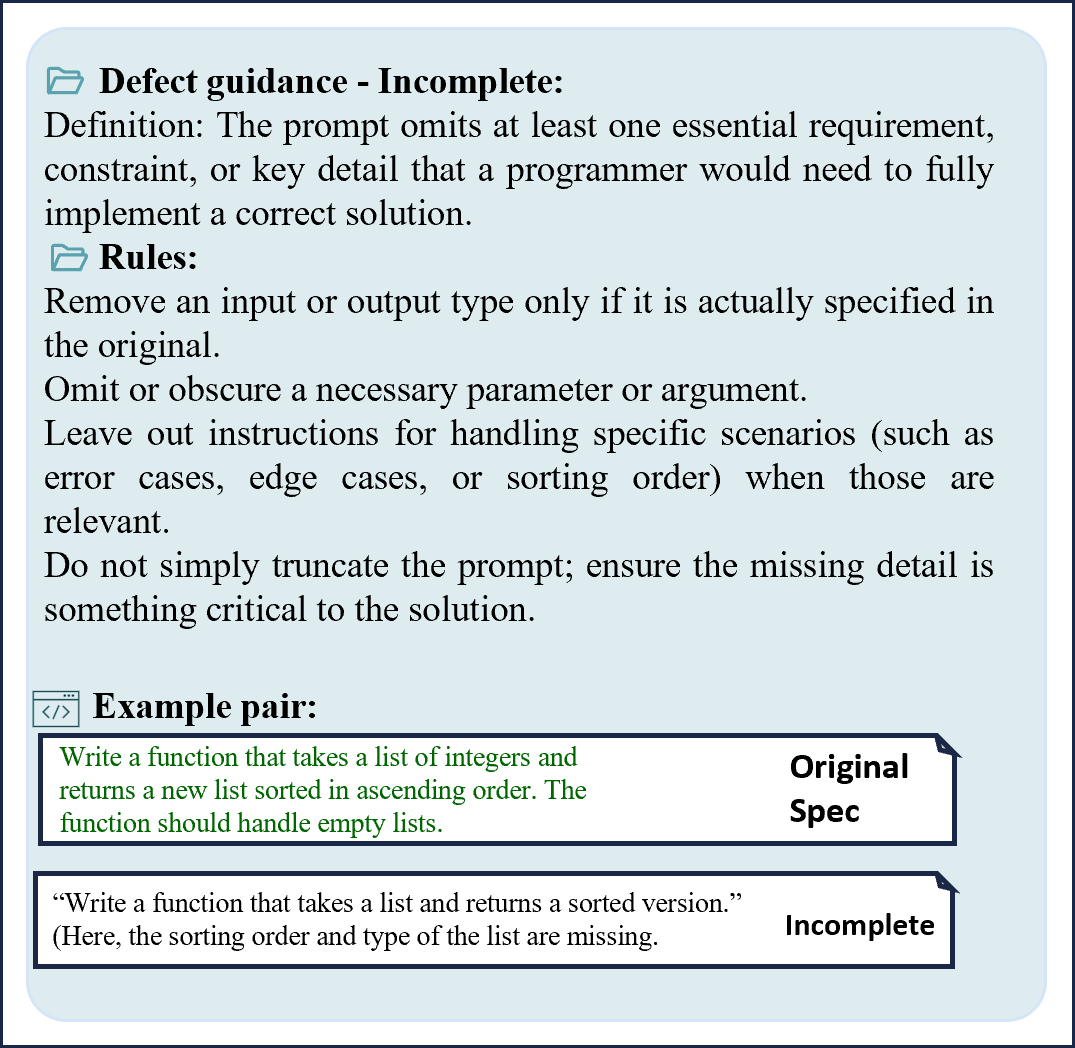}
    \caption{Prompt for incompleteness defect injection.}
    \label{fig:defect_injection_incomplete}
\end{figure}

\begin{figure}[ht]
    \centering
    \includegraphics[width=1\linewidth]{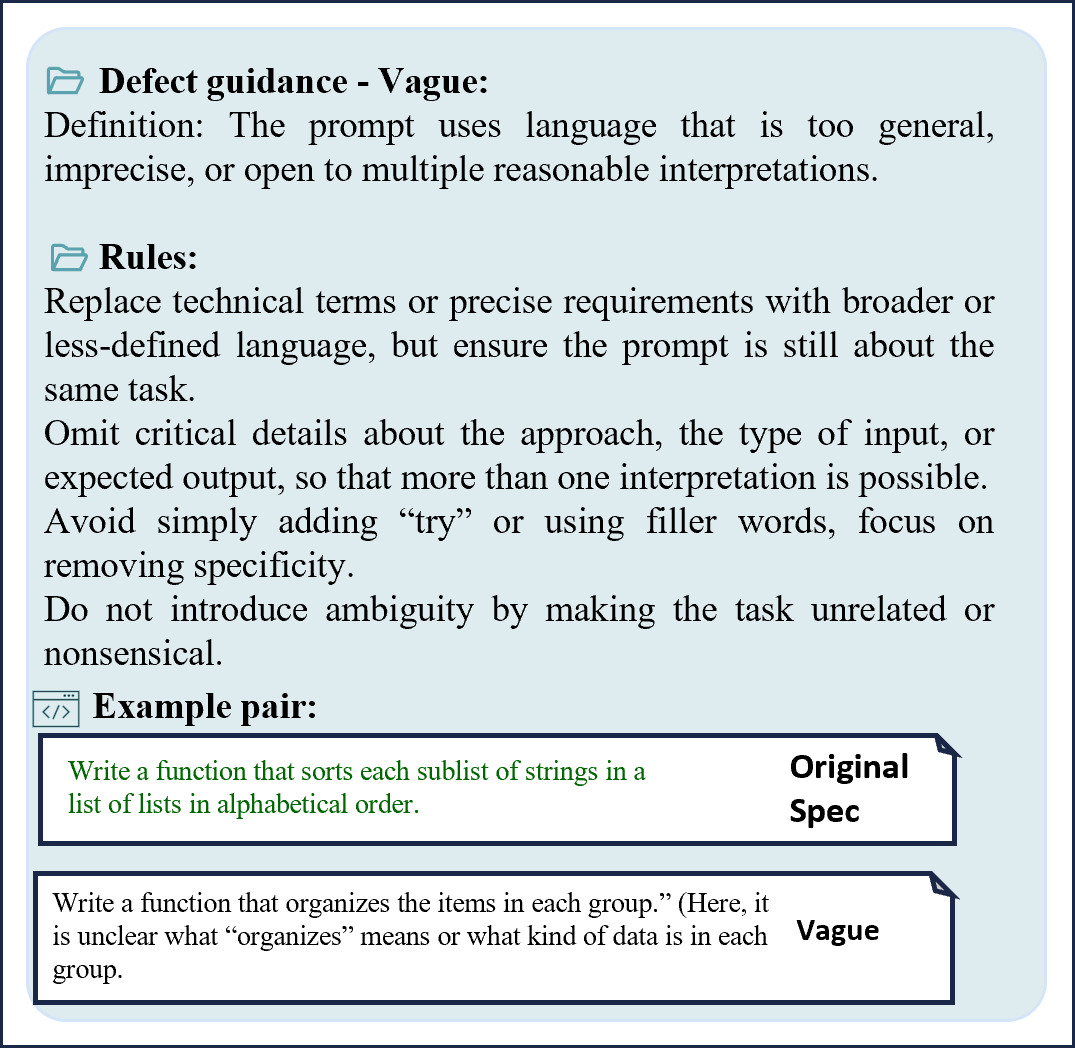}
    \caption{Prompt for vagueness defect injection}
    \label{fig:defect_injection_vague}
\end{figure}

Fig.~\ref{fig:main_defect_injection} shows the base prompt used for all defect injections, including general instructions and a software-domain example. The defect-specific prompts can be found in Fig.~\ref{fig:defect_injection_contradictory} ,Fig.~\ref{fig:defect_injection_incomplete} and Fig.~\ref{fig:defect_injection_vague}

\clearpage

\clearpage
\onecolumn
\section*{Appendix III: Detail taxonomy of defect specification and the framework's performance}
\label{app:3_performance_taxonomy}

\begin{table*}[]
\centering
\caption{Performance comparison of Spec-level and Hybrid Repair. Results show Pass@1 across difficulty levels (m=1, m=2, m=3).}
\label{tab:performance_parameter}
\small
\begin{tabular}{lcccccccc}
\toprule
\multirow{3}{*}{Model} & \multirow{3}{*}{Method} & \multirow{3}{*}{Defect Type} & \multicolumn{6}{c}{Benchmark} \\
\cmidrule(lr){4-9}
 & & & \multicolumn{3}{c}{VerilogEval-defect} & \multicolumn{3}{c}{ComplexVDB-defect} \\
\cmidrule(lr){4-6} \cmidrule(lr){7-9}
 & & & m=1 & m=2 & m=3 & m=1 & m=2 & m=3 \\
\midrule
\multirow{8}{*}{DeepSeek-v4-flash} & \multirow{4}{*}{Spec-level} & All & 0.440 & 0.429 & 0.415 & 0.255 & 0.238 & 0.236 \\
 & & Contradictory & 0.556 & 0.565 & 0.561 & 0.238 & 0.219 & 0.217 \\
 & & Vague & 0.318 & 0.292 & 0.279 & 0.277 & 0.266 & 0.264 \\
 & & Incomplete & 0.447 & 0.429 & 0.406 & 0.249 & 0.228 & 0.228 \\
\cmidrule(lr){2-9}
 & \multirow{4}{*}{Hybrid Repair} & All & 0.500 & 0.491 & 0.473 & 0.384 & 0.356 & 0.354 \\
 & & Contradictory & 0.595 & 0.607 & 0.597 & 0.356 & 0.331 & 0.331 \\
 & & Vague & 0.379 & 0.351 & 0.337 & 0.387 & 0.359 & 0.356 \\
 & & Incomplete & 0.527 & 0.514 & 0.487 & 0.409 & 0.379 & 0.376 \\
\midrule
\multirow{8}{*}{GPT-5.4-nano} & \multirow{4}{*}{Spec-level} & All & 0.354 & 0.358 & 0.360 & 0.258 & 0.245 & 0.240 \\
 & & Contradictory & 0.449 & 0.479 & 0.485 & 0.268 & 0.262 & 0.255 \\
 & & Vague & 0.219 & 0.214 & 0.214 & 0.225 & 0.211 & 0.204 \\
 & & Incomplete & 0.394 & 0.381 & 0.381 & 0.283 & 0.260 & 0.260 \\
\cmidrule(lr){2-9}
 & \multirow{4}{*}{Hybrid Repair} & All & 0.443 & 0.444 & 0.444 & 0.368 & 0.336 & 0.329 \\
 & & Contradictory & 0.547 & 0.566 & 0.574 & 0.412 & 0.391 & 0.377 \\
 & & Vague & 0.290 & 0.285 & 0.285 & 0.303 & 0.286 & 0.278 \\
 & & Incomplete & 0.493 & 0.481 & 0.474 & 0.390 & 0.332 & 0.332 \\
\bottomrule
\end{tabular}
\end{table*}

\textbf{Effect of Maximum Inconsistency Pairs (m)}

Table~\ref{tab:performance_parameter} presents an ablation study on the maximum number of inconsistency pairs ($m$) extracted during Spec-Level Repair. We observe a trade-off that varies by defect type.

\textit{Contradiction defects benefit from larger $m$.} For contradiction-type defects, VClare Full achieves its highest performance at $m=2$ on DeepSeek-V4-Flash (60.7\% on VerilogEval-Defect) and continues improving up to $m=3$ on GPT-4 (57.4\%). This indicates that the LLM can effectively leverage additional mined inconsistency pairs when the specification contains explicit conflicting statements.

\textit{Vague and incomplete defects degrade with larger $m$.} For vague and incomplete defects, performance consistently decreases as $m$ increases from 1 to 3. On DeepSeek-V4-Flash with vague defects, Pass@1 drops from 37.9\% ($m=1$) to 33.7\% ($m=3$). Similarly, incomplete defects decline from 52.7\% to 48.7\%.

\textit{Interpretation: Localization vs. Repair.} Our analysis reveals the underlying cause of this divergence. As $m$ increases, the LLM successfully identifies more potential defect locations—including the true injected defect. However, for vague and incomplete defects, the LLM lacks sufficient information to determine the \emph{correct} repair action. Unlike contradiction defects where the specification provides both the correct and incorrect statements (enabling the LLM to choose), vague and incomplete defects offer no such ground truth. Consequently, the LLM introduces spurious or incorrect modifications, harming overall performance. This suggests that our automative fix of LLM is creating more harm than good, and indicates that it is best practice for future system to call for more information from human engineer.

\textit{Implication for automated repair.} In a fully automated setting (without human arbitration), $m=1$ represents the safest configuration, minimizing the risk of spurious edits. With human-in-the-loop arbitration, larger $m$ values become viable because engineers can validate both defect localization and proposed repairs. This suggests that practical deployment should either (a) limit automated repair to $m=1$, or (b) incorporate lightweight human validation of mined pairs before repair.
\twocolumn

\end{document}